\documentclass[aps,pra,twocolumn,groupedaddress,amsmath,amssymb]{revtex4-1}
\usepackage{graphicx}% Include figure files
\usepackage{epsfig}
\usepackage{dcolumn}% Align table columns on decimal point
\usepackage{bm}% bold math
\usepackage{amssymb}
\usepackage{bm}
\usepackage{amsmath}
\usepackage{braket}
\begin{document}
\title{Bound state in the continuum by spatially separated ensembles of atoms in a coupled-cavity array}
\author{P. T. Fong and C. K. Law}
\affiliation{
   Department of Physics and Institute of Theoretical Physics, The Chinese University of Hong Kong, Shatin, Hong Kong Special Administrative Region, China}
\date{\today}
\begin{abstract}

We present an analytic solution of bound states in the continuum (BICs) for photons and atoms 
in a one-dimensional coupled-cavity array. These bound states are formed by 
two ensembles of two-level atoms confined in separated cavities of the array. We show that in the regime where the
photon hopping rate between neighboring cavities is high compared with the collective Rabi frequency, 
the BIC corresponds to a nonradiating collective atomic state in which the two ensembles of atoms are strongly entangled. 
On the other hand in the low photon hopping rate regime, the BIC behaves as a quantum cavity in which photons can be 
trapped between the two ensembles of atoms. 
 
 \end{abstract}
\maketitle
\section{Introduction}
Bound states in the continuum (BICs) are spatially localized states with energies embedded in the continuous spectrum of a system. 
The first example of a BIC for a single quantum particle was 
proposed by von Neumann and Wigner\cite{Wigner}. Later, Stillinger and Herrick explored the possibilities of BICs in two-electron systems \cite{Stillinger1, Stillinger2}. A formal theory of BICs based on interfering resonances was developed
by Friedrich and Wintgen \cite{Wintgen}. It is now known that BIC is a wave phenomenon that can be realized in a variety of physical wave systems \cite{review}. In particular, BICs have been observed 
in various (classical) photonic systems such as waveguides \cite{Dreisow,bicwaveguide1,bicwaveguide2}, photonic crystals \cite{BICreview,Gansch} and photonic Lieb lattices \cite{bicphotonlattice1,bicphotonlattice2}. There also exist surface bound states in the continuum 
in photonic structures \cite{surface1,surface2}. Since photonic BICs
correspond to a high-quality confinement of light at specific frequencies, 
they have useful applications in optical devices such as filters \cite{Marinica} 
and lasers \cite{laser}. 

Although the concept of BICs was originally proposed as a solution of the Schr\"odinger equation, there have been relatively few 
studies of BICs in quantum systems. A difficulty is a lack of quantum systems
processing such a kind of bound states in general. However, recent investigations in waveguide QED have found interesting examples of 
BICs for dressed photon-atom systems \cite{kim1,Moreno,kim2,Kocabas,Facchi,longhi,zhou,decoherencefree}. 
Waveguide QED generally refers to photon-atom interactions in one-dimensional photonic structures, such as superconducting qubits in transmission lines  \cite{Blais,Wallraff}, and trapped atoms in a nanophotonic waveguide \cite{Goban,Lodahl}. Since atoms can be strongly coupled to tightly confined waveguide modes, waveguide QED has been a useful platform for studying photon-photon interaction \cite{Roy} and quantum information processing \cite{Blais2,kimble1,baranger}. In particular, a coupled-cavity array can have energy bands 
equivalent to that of the tight-binding model, and interesting effects such as  photon scattering \cite{zhou2,Kocabas2},  single-photon transport \cite{transport1, transport2, transport3}, 
and photon-atom bound states outside the continuum \cite{rabl1,tshi} have been discussed in literature.

We note that BICs comprising a single excitation (or photon) have 
been studied theoretically 
in a coupled-cavity array  \cite{longhi,zhou,decoherencefree}. Such BICs are typically
formed by two separated two-level atoms, such that a single photon at a resonance 
frequency can be perfectly trapped in the space between the two atoms. In this regard,
the two atoms behave as a quantum cavity for a single photon.

In this paper we show that BICs comprising multiple excitations or photons can be formed by two spatially separated
ensembles of two-level atoms in a coupled-cavity array. This extends the previous studies \cite{longhi,zhou,decoherencefree}
to situations of multiple photons. We note that the problem of two-particle BICs has been studied
in various Bose-Hubbard models \cite{BH1,BH2,BH3} and interesting interference effects due to quantum statistics of particles have been 
reported \cite{Crespi}. Here we present a type of multi-particle BICs formed by two distinct constituents, namely 
photons and atoms in waveguide QED, and the atomic degree of freedom is described by collective spin variables. 
As we shall see below, our BICs exhibit various features 
depending on the photon hopping rate between neighboring cavities and the photon-atom interaction strength. 
Apart from the quantum-cavity effect for multiple photons, there exists a subradiance regime where excitations are mostly stored in 
a collective atomic state that does not radiate. We note that subradiance is a collective quantum 
phenomenon that has attracted research interest recently in experimental  \cite{sub1,sub2,sub3} 
and theoretical studies \cite{sub4a,sub4,sub5,sub6,sub7}. Our paper provides an example of subradiance by two 
ensembles of atoms.

Our paper is organized as follows. We first describe the system and the corresponding Hamiltonian in Sec. II. 
Then by making use of a decoupling condition, we obtain an exact analytic solution of BICs with a general 
excitation number in Sec. III. In Sec. IV, we focus on the properties of BICs in 
a triple-cavity system. Such a system allows us to analyze the structures of BICs as well as their formation in detail. 
In particular, we indicate how subradiant states can be formed dynamically by free evolution in the subradiance regime. 
We also provide a linear theory of the system in the quantum-cavity regime and determine how atomic decoherence 
affects the storage time of photons. The conclusions and remarks are given in Sec. V.

\section{The Model}

We consider a one-dimensional coupled-cavity array formed by $N+1$ (where $N>1$) 
cavities, in which the leftmost and rightmost cavities each contains $M$ identical 
two-level atoms (Fig. 1). We assume that each cavity mode has the same resonance 
frequency $\omega_c$ which is close to the atomic transition frequency $\omega_A$.
The full Hamiltonian of the system is given by ${\cal H} = H_1 + H_2$ 
where ($\hbar=1$),
\begin{eqnarray}
H_1 & = & \omega_c(a^{\dagger}_La_L+a^{\dagger}_Ra_R) + \omega_A(J_L^z+J_R^z) +  \sum_{n=1}^{N-1}\omega_cb^{\dagger}_nb_n  \nonumber\\ 
&& + g( a^{\dagger}_{L} J^-_L + a^{\dagger}_{R} J^-_R  + a_{L} J^+_L + a_{R} J^+_R) \nonumber \\
&& +\lambda  ( a^{\dagger}_{L}  b_1+a^{\dagger}_{R} b_{N-1}+ a_{L}  b_1^\dag + a_R b_{N-1}^\dag ) \nonumber \\
&& +\lambda\sum_{n=1}^{N-2}(b^{\dagger}_nb_{n+1}+ b_{n+1}^\dag b_n ), \\
H_2 & = & \int_{0}^{\infty}d \omega \omega [ c^{\dagger}_L(\omega)c_L(\omega)+ c^{\dagger}_R(\omega)c_R(\omega)] \nonumber\\
&& + \int_{0}^{\infty}d \omega [\eta_L(\omega)c^{\dagger}_L(\omega)a_L+\eta_R(\omega)c^{\dagger}_R(\omega)a_{R}+ {\rm H.c.}]. \nonumber \\
\end{eqnarray}
Here $H_1$ describes the photon-atom interaction (second line) 
and tight binding type coupling between neighboring cavities (third and fourth line), 
and $H_2$ describes outside field modes and their interaction with the fields in 
the end cavities. Specifically, $a_{L}$ and $a_{R}$ are 
annihilation operators for the leftmost and rightmost cavity modes  respectively, $b_n $ $(n=1,2,..., N-1)$ 
are annihilation operators associated with the cavities in the middle, and $\lambda$ (assumed real) 
is the coupling constant between adjacent cavities \cite{pt}. In our model, 
we assume that all atoms in the respective cavities experience the same photon-atom coupling strength $g$, and so atoms are treated collectively. For the $M$ atoms in the leftmost cavity, 
we define,
\begin{eqnarray}
&& J^z_{L}=\frac{1}{2}\sum^{M}_{m=1}(\ket{1}_m\bra{1}-\ket{0}_m\bra{0}),  \\
&& J^+_{L}=\sum^{M}_{m=1}\ket{1}_m\bra{0} =  (J_{L}^{-})^{\dag}.
\end{eqnarray}
Here $|1\rangle_m$ and $|0\rangle_m $ denote the excited and ground state of the $m$th atom. 
For the atoms in the rightmost cavity, $J_R^z$ and $J_R^+$ are defined similarly but with
the summation index $m$ taken from $M+1$ to $2M$. Finally for the $H_2$, we have used $c_L(\omega)$ and $c_R(\omega)$, respectively, for the annihilation operators of continuous field modes at the frequency $\omega$ outside the leftmost and rightmost cavities, 
and $\eta_L(\omega)$ and $\eta_R(\omega)$ are coupling strengths \cite{note1}. 

\begin{figure}
\includegraphics[width=3.3 in]{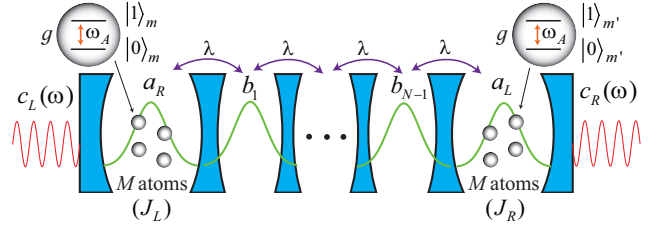}  
\caption{A schematic drawing of the one-dimensional coupled-cavity array. The leftmost and rightmost cavities 
each contains $M$ identical two-level atoms. The coupling parameters
and operators defined in the text are indicated.}
\end{figure}

The Hamiltonian $H_1$ can be simplified by using the normal modes of the
cavity chain (excluding the two end cavities). The annihilation operators associated with
such normal modes are given by
\begin{eqnarray}
B_{k}=\sqrt{\frac{2}{N}}\sum_{n=1}^{N-1}b_n\sin{\frac{kn\pi}{N}}, 
\end{eqnarray} 
with $k=1,2,...,N-1$. Such operators satisfy $[B_{k}, B_{k'}^\dag] = \delta_{kk'}$. In this way, 
$H_1$ reads,
\begin{eqnarray}
H_1 & = & \omega_c(a^{\dagger}_La_L+a^{\dagger}_Ra_R) + \omega_A(J_L^z+J_R^z) 
\nonumber\\
&& +\sum_{k=1}^{N-1}\Omega_{k}B^{\dagger}_{k}B_{k} + [ a^{\dagger}_{L} (gJ^-_L + \sum_{k=1}^{N-1} \lambda_{k}^L B_k) + {\rm H.c.} ] \nonumber \\
&& + [a^{\dagger}_{R} (gJ^-_R + \sum_{k=1}^{N-1} \lambda_{k}^RB_k) + {\rm H.c.}]  
\end{eqnarray} 
where
\begin{eqnarray}
&&\lambda_{k}^L=\lambda\sqrt{\frac{2}{N}}\sin{\frac{k\pi}{N}}, \\
&&\lambda_{k}^R=\lambda\sqrt{\frac{2}{N}}\sin{\frac{k(N-1)\pi}{N}} = (-1)^{(k+1)} \lambda_{k}^L,\\
&&\Omega_k=\omega_c+ 2 \lambda \cos{\frac{k\pi}{N}}
\end{eqnarray} 
are defined. For later purposes, we introduce the operator
\begin{eqnarray}
{\cal N} = a^{\dagger}_L a_L   +  a^{\dagger}_R a_R+  J_{L}^z + J_{R}^z + M  + \sum_{k=1}^{N-1} B^{\dagger}_{k}B_{k} \nonumber \\
\end{eqnarray}
which corresponds to the total number of excitations stored in the cavity array and atoms.

\section{Solutions of the BIC}

Owing to the coupling to the continuum, the full Hamiltonian $\cal H$ 
has a continuous energy spectrum with energy $E \ge - M \omega_A$. A BIC represents a localized eigenstate with 
its energy  embedded in the continuous energy spectrum. Here we 
adopt a strict localization condition that all excitations are trapped inside
the atoms and the cavity array, while all the field modes outside [described by $c_L (\omega)$ and  $c_R (\omega)$] 
are in the vacuum state $|vac \rangle_{\rm out}$. For a given excitation number $K$, the BIC takes the form:
\begin{equation}
|\Psi_{BIC}^{(K)} \rangle = | \beta_K \rangle |vac\rangle_{\rm out}
\end{equation}
where $| \beta_K \rangle$ is a trapped state describing the state of atoms and photons inside the cavity array.
Specifically,  $| \beta_K \rangle$ is a common eigenvector of $H_1$ and ${\cal N}$:
\begin{eqnarray}
&& H_1 | \beta_K \rangle = E_b | \beta_K \rangle,  \\
&& {\cal N}|\beta_K\rangle = K |\beta_K\rangle,
\end{eqnarray}
where $K$ is an integer, and $E_b$ is the energy eigenvalue.

In this paper we present a solution of  $|\beta_K\rangle$  with a given excitation number $K$ in the form:
\begin{equation}
|\beta_K\rangle=\sum^{K}_{m=0}\sum^{K-m}_{n=0}\alpha_{mn}B_q^{\dag m} (J_L^{+})^n  (J_R^{+})^{K-m-n} \ket{\phi}.
\end{equation}
Here $\alpha_{mn}$ are some coefficients, and 
$\ket {\phi}$ is the state in which all the field modes in the cavity array
are in the vacuum state and all the atoms are in their $|0\rangle$ state. In writing Eq. (14), we have assumed that 
there exists a mode described by $B_q$ which has the frequency $\Omega_q =\omega_A$. For example, such a mode occurs 
when $q=N/2$ for an even $N$ and $\omega_c=\omega_A$. The effect of a nonzero detuning ($\Omega_q \ne \omega_A$) will be
discussed in the next section. 

Noting that $|\beta_K\rangle$ defined by Eq. (14) has zero photons in the end cavities,  $|\beta_K\rangle$
is an eigenvector of $H_1$ with the energy eigenvalue $E_b=(K-M)\omega_A$ if the following conditions are satisfied:
\begin{eqnarray}
&&(gJ^-_{L}+\lambda_{q}^LB_{q})|\beta_K\rangle=0, \\
&&(g J^-_R+\lambda_{q}^RB_{q})|\beta_K\rangle=0.
\end{eqnarray}
Physically, these two conditions can be interpreted as a destructive interference between the
photon emission by atoms [described by $ga_{\mu}^\dag J^-_{\mu}$ in Eq. (6)] and the photon tunneling 
[described by $\lambda_{q}^{\mu} a_{\mu}^\dag B_q$ in Eq. (6)]. Together with the vacuum field outside, the conditions (15) and (16) 
ensure that the fields in the end cavities remain in the vacuum state. Since the end cavity modes are never
excited, no energy can escape to the continuum, and hence all the excitations are trapped
in the atoms and the $B_q$ mode. Unlike BICs based on destructive interference 
in classical electromagnetic wave systems \cite{review,Yang},
here the interference corresponds to the inhibition of quantum transitions in a finite
dimensional Hilbert space.

The conditions Eqs. (15) and (16) require $M \ge K$. To determine the explicit form of $|\beta_K\rangle$, we rewrite Eq. (14) as
\begin{eqnarray}
|\beta_K\rangle = \sum^{K}_{m=0}\sum^{K-m}_{n=0}c_{m,n}\ket{m,n,K-m-n} |0_L,0_R \rangle  \ \ 
\end{eqnarray}
where 
$|0_L, 0_R \rangle$ denotes the vacuum state of the two end cavities,
$c_{m,n}$'s are coefficients, and $\ket{m,n,K-m-n}$ is a common eigenvector of $B_q^\dag B_q$, $J_\mu ^2,$ and $J_\mu ^z$ ($\mu = L, R$)
such that,
\begin{eqnarray}
&& B_q^\dag B_q \ket{m,n,r} = m \ket{m,n,r}, \\
&& J_{L}^z \ket{m,n,r} = (n-\frac {M} {2}) \ket{m,n,r}, \\
&& J_{R}^z \ket{m,n,r} = (r-\frac {M} {2}) \ket{m,n,r}.
\end{eqnarray}
In words, $m$ is the photon number in the $B_q$ mode, and $n$ and $r$ are the number of excited atoms in the left and right end 
cavities, respectively. 

By using the conditions (15) and (16), we obtain the recursive relations:
\begin{eqnarray}
&& c_{m+1,n-1}=-\frac{g}{\lambda^L_q}\frac{\sqrt{n(M-n+1)}}{\sqrt{m+1}}c_{m,n}, \\
&& c_{m+1,n}=-\frac{g}{\lambda^R_q}
 \frac{\sqrt{(K-m-n)(M-K+m+n+1)}}{\sqrt{m+1}} c_{m,n}. \nonumber \\
\end{eqnarray}
This leads to the solution of $c_{m,n}$ given by,
\begin{eqnarray}
c_{m,n}&=&(\chi)^{m}(\frac{\lambda_q^R}{\lambda_q^L})^n
\sqrt{\frac{(M-K+n+m)!}{(K-n-m)! m!}}\nonumber\\
& & \times \sqrt{\frac{(M-n)!}{M!}} \sqrt{\frac{K!}{(M-K)!n!}} c_{0,0},
\end{eqnarray}
where the parameter $\chi$ (assumed positive) is defined by:
\begin{equation}
\chi = -g/\lambda_q^R = (-1)^q g/\lambda_q^L 
\end{equation}
and the value of $c_{0,0}$ is determined by the normalization condition.
Note that $(\lambda_q^R/\lambda_q^L) = (-1)^{q+1}$ depends on the integer mode index $q$.  

Equation (23) is a main result of this paper. For a given $M$ and $K$, the structure of the BIC is controlled by
the parameter $\chi$ since the $m$-photon amplitude scales with $\chi^m$. In the following we discuss the 
features in $\chi \gg 1$ and $\chi \ll 1$ regimes.

\subsection{Subradiance regime }

In the $\chi \ll 1$ limit, the zero photon amplitudes $c_{0,n}$ contribute most to 
the BIC. Therefore the BIC is mainly formed by excited atoms in a collective state 
that does not radiate, i.e., a subradiant state. By keeping $c_{0,n}$ terms only, we have
\begin{eqnarray}
|\beta_K\rangle \approx \sum^{K-m}_{n=0}c_{0,n}\ket{0,n,K-n} |0_L,0_R \rangle.
\end{eqnarray}
In particular,
the state with the excitation number $K=M$ is a maximally entangled state in which the atomic excitations
in the two ensembles are perfectly corrected,
 \begin{eqnarray}
|\beta_M\rangle \approx  \frac{1}{\sqrt{M+1}}\sum^{M}_{n=0}(\pm 1)^{n}\ket{0,n,M-n} |0_L,0_R \rangle, \ \
\end{eqnarray}
where $+1$ and $-1$ are for odd and even $q$, respectively. As we shall see in the next section
the right hand side of Eq. (26) is a singlet state.

Note that the specific requirement of the smallness of $\chi$ for the subradiant regime 
depends on the excitation number $K$. To ensure that the zero photon amplitudes $c_{0,n}$ 
are dominant, we require $|c_{m+1,n}| / |c_{m,n}| \ll 1$. By Eq. (22), it can be shown that 
$|c_{m+1,n}| / |c_{m,n}| < \chi \sqrt{(1+M-r)r}$ with $r$ defined in Eq. (20). 
Noting that $r$ is bounded by $0\le r \le K-n-m$, $|c_{m+1,n}| / |c_{m,n}| \ll 1$ would need $\chi (M+1) /2 \ll 1 $ 
for $K=M$, and  $\chi \sqrt M \ll 1 $ for $K \ll M$.

\subsection{Quantum-cavity regime}

In the regime where $\chi \gg 1$, $c_{K,0}$ in Eq. (23) is much larger than all other amplitudes.
Hence $|\beta_K\rangle$ is approximately a Fock state:
\begin{eqnarray}
|\beta_K\rangle  \approx  |K,0,0 \rangle |0_L,0_R \rangle,
\end{eqnarray}
which indicates that almost all excitations are stored as photons in the 
the field mode $B_q$. In other words, the two ensembles of atoms 
effectively form a quantum cavity for multiple photons. 

In Fig. 2, we illustrate the transition between two regimes by showing the expectation values of photon number 
in the mode $B_q$ and the atomic excitation number for the state $|\beta_M\rangle$ as a function of $\chi$. For the case with the excitation number 
$K=M=2$ used in the figure, $\chi > 5$ is sufficient to have more than $90\%$ of photonic excitations. On the other 
hand $\chi < 0.5$ would give more than $90\%$ of excitation stored in the atoms.

\begin{figure}
\includegraphics[width=3 in]{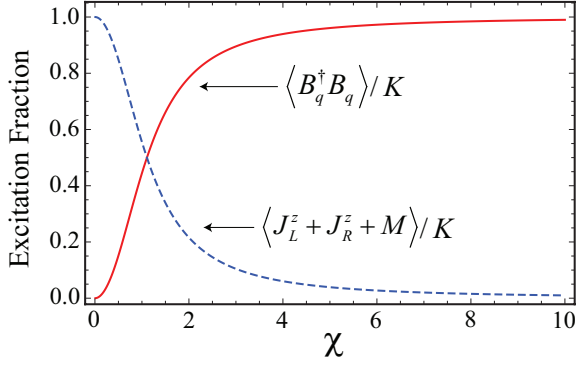}  
\caption{The average number of photons (solid line) and excited atoms (dashed line) 
normalized by the total number of excitation as a function of $\chi$ for the BIC with
$M=K=2$.}
\end{figure}

\section{Triple-cavity configuration}

In this section we examine the BICs in a triple-cavity configuration 
($N=2$ in Fig. 1) in which there is only one normal mode $B_1=b_1$ for the cavity in the middle. 
Such a configuration is conceptually simpler and it provides useful physical insights about 
the formation of BICs.  

\subsection{Effective Hamiltonian in the subradiant regime}

First we introduce the following operators:
\begin{eqnarray}
&& a_{-} = \frac{a_L - a_R}{\sqrt{2}}, \\
&& \nu_{\pm} = \frac{a_L+a_R \pm \sqrt 2 b_1}{2},
\end{eqnarray}
which satisfy $[a_-, a_-^\dag] = [\nu_+ , \nu_+ ^\dag] =  [\nu_- , \nu_- ^\dag]=1$, and 
$[a_-,\nu_{\pm}^\dag]=[\nu_+, \nu_-^\dag]=0$. These operators correspond to creation and
annihilation operators associated with the normal modes of the three coupled cavities.
Accordingly, the Hamiltonian $H_1$ in Eq. (1) can be rewritten as,
\begin{eqnarray}
H_1 &=& \omega_A S^z + \omega_c a^{\dagger}_-a_- + \frac{g}{\sqrt{2}}(S^- a^{\dagger}_- + {\rm H.c.}) \nonumber\\
& +& (\omega_c + \sqrt{2}\lambda)\nu^{\dagger}_{+}\nu_{+} +(\omega_c - \sqrt{2}\lambda)\nu^{\dagger}_{-}\nu_{-} \nonumber\\
& +& \frac{g}{2} \big[ (J^{-}_L - \tilde{J}^{-}_R) (\nu^{\dagger}_+ + \nu^{\dagger}_-) + {\rm H.c.} \big].
\end{eqnarray}
Here we have defined $S^z = J^{z}_L + \tilde{J}^{z}_R$ and $S^{\pm} = J^{\pm}_L + \tilde{J}^{\pm}_R$, 
with $\tilde{J}^{\pm}_R = -J^{\pm}_R$ and $\tilde{J}^z_R = J^z_R$.  
Note that $S^i$ ($i=z,\pm$) obey the commutation relations of angular momentum operators: $[ S^+,S^- ] = 2S^z$ and 
$[ S^z, S^{\pm} ]=\pm S^{\pm}$. Since there are $M$ identical two-level atoms in each
of the end cavity, $S^i$'s are equivalent to the addition of angular momentum operators for 
two spin-$M/2$ systems. 
 
Let $|s, m_s \rangle$ be a common eigenvector of $S^2$ and $S^z$, then
\begin{eqnarray}
&& S^2|s, m_s \rangle = s(s+1)|s, m_s \rangle,  \\
&& S^z|s, m_s \rangle = m_s |s, m_s \rangle ,
\end{eqnarray}
where the allowed quantum numbers are $s = 0,1,2,...,M$ and $m_s = -s,-s+1,...,s$. 

Now returning to the Hamiltonian (30) and assuming $\omega_c \approx \omega_A$, we see that 
the atoms and $\nu_{\pm}$ modes are essentially uncoupled in the subradiance regime because
the detunings $\pm \sqrt 2 \lambda$ are much larger than $g$ in magnitude. Therefore we may
neglect the far off resonant interaction terms in the last line of the Eq. (30). Such an approximation 
is a type of rotating wave approximation when viewing the system in the interaction picture. 
By keeping resonant terms of Eq. (30) (i.e., discarding  $\nu_{\pm}$ modes) ,
atoms and the $a_-$ mode photons are coupled via the effective Hamiltonian
\begin{eqnarray}
H_1' = \omega_A S^z + \omega_c a^{\dagger}_-a_- + \frac{g}{\sqrt{2}}(S^- a^{\dagger}_- + {\rm H.c.}),
\end{eqnarray}
which is the Hamiltonian of the Tavis-Cummings model \cite{TC,Fink}.
Since $[S^2, H_1']=0$, the quantum number $s$ is a constant of motion.

According to the effective Hamiltonian $H_1'$, atoms in the collective state
$|s, m_s = -s \rangle$ are nonradiating, since they cannot emit a photon to 
the $a_-$ mode by lowering the quantum number $m_s$. This is connected to 
the BIC in the previous section, where the $|\beta_K \rangle$ given in Eq. (25)
can be rewritten as: 
\begin{equation}
|\beta_{M-s} \rangle \approx |s, m_s = -s \rangle |0_b \rangle |0_L,0_R \rangle,
\end{equation}
with $|0_b \rangle$ being the vacuum field state in the $b_1$ mode. 
For example, Eq. (26) corresponds to the singlet state with $s=0$. 
A general derivation of Eq. (34) can be obtained by using Eq. (25) with $c_{0,n}$ 
obtained in Eq. (23) and Clebsch-Gordan coefficients.

\subsection{Evolution to subradiant states}

Consider the system in the subradiance regime with atoms initially prepared in a superposition of $|s, m_s \rangle$
states and no photon in the cavities, the initial state is given by,
\begin{equation}
|\psi (0) \rangle = \sum_{s=0}^M \sum_{m_s = -s}^s C_{s,m_s} |s,m_s \rangle |0_b \rangle |0_L,0_R \rangle,
\end{equation} 
where $C_{s,m_s}$ are some coefficients. As the system evolves,
atoms would be de-excited by emitting 
a photon into the $a_-$ mode according to $H_1'$, and hence the quantum number $m_s$ is lowered by 1. 
Furthermore, since the $a_-$ mode is damped because of its coupling to 
the outside modes [governed by $H_2$ in Eq. (2)], de-excitation of atoms will continue until
the quantum number $m_s$ reaches the lowest possible value $m_s = -s$.  
Therefore the system would eventually be trapped as a mixture of  $|s,-s \rangle$ for different $s$'s.

In terms of the reduced density matrix $\rho$ obtained by taking the trace over the outside field modes, 
the final $\rho$ would be a mixed state:
\begin{equation}
\rho \approx \sum_{s=0}^M  p_s  |s,-s \rangle \langle s, -s| \otimes |0\rangle_{TT} \langle 0|,
\end{equation} 
where  
$p_s =\sum_{m_s=-s}^s|C_{s,m_{s}}|^2$ and $|0\rangle_T \equiv  |0_b \rangle |0_L,0_R \rangle$ for brevity,

To verify the above result, we calculate the time evolution of $\rho$ by the master equation method. 
Assuming that the coupling strengths $\eta_L (\omega)$ and 
$\eta_R(\omega)$ in Eq. (2) are frequency independent,
the outside field modes are equivalent to Markovian oscillator baths at zero temperature. This leads to 
the usual Markovian master equation,
\begin{eqnarray}
\dot \rho = -i \big[ H_1, \rho \big]+\sum_{\mu=L,R}\frac{\gamma_c}{2}{\cal D}[a_\mu] \rho,
\end{eqnarray}
where $H_1$ is the original Hamiltonian in Eq. (1) without using the approximation in the 
previous subsection, and the super-operator ${\cal D}$ is defined by: 
\begin{equation}
{\cal D}[X] \rho =(2 X \rho X^{\dagger} - X^{\dagger} X \rho - \rho X^{\dagger} X).
\end{equation}
In writing Eq. (37), we have used $\eta_L (\omega)= \eta_R (\omega)= \sqrt {\gamma_c/2\pi}$, with
$\gamma_c$ being the leakage rate of the end cavities. 
Note that decoherence effects on atoms
due to interactions with noncavity modes have been omitted in the master equation. This is justified
as long as the relevant relaxation rates of atoms are sufficiently small so that the steady state of $\rho$ defined by 
Eq. (37) can be established before decoherence effects become significant.  

We have solved the master equation numerically for the $M=2$ case, and the numerical steady
state agrees well with the approximation given in Eq. (36). Specifically, we consider
the two atoms in the left (right) cavity are initially prepared in the excited (ground) states, 
and all the field modes are in the vacuum state. In this case, $s=0,1,2$ are allowed quantum
numbers. 

In Fig. 3 we illustrate the numerical results by showing the time dependence of the probability in 
the trapped state $|\beta_i \rangle$, which is $P_i = \langle \beta_i |\rho |\beta_i \rangle $, obtained
from the master equation.
We see that $P_2$ is a constant because the initial system with an excitation number $K=2$
has a partial overlap with trapped state
$|\beta_2 \rangle$, and it will remain a constant. As the system evolves,
$P_0$ and $P_1$ increase from zero and become steady, and so the system is trapped in the lower states 
$|\beta_1 \rangle$ and $|\beta_0 \rangle$.
Note that $|\beta_0 \rangle$ is just the trivial ground state of the system. 
For the parameters used in Fig. 3, we have verified that $|\beta_2 \rangle$ and $|\beta_1 \rangle$ are well   
approximated by Eq. (34) and the steady value of $P_i$ agrees well with $p_{M-i}$ defined after Eq. (36) within 
$1\%$ discrepancy.

\begin{figure}
\includegraphics[width=3 in]{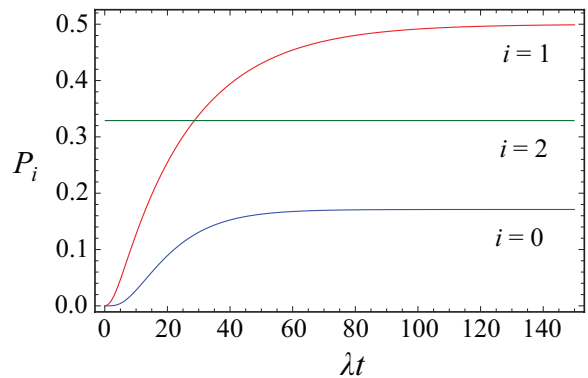}  
\caption{Occupation probabilities of the states $\ket{\beta_i}$ for $i=0,1,2$ as a function of dimensionless time $\lambda t$. 
The evolution is obtained by solving the master equation (37) numerically with the  
parameters $M=2$, $N=2$, $g/\lambda=0.1$, $\gamma_c/ \lambda = 1$ and the initial state has an excitation number $K=2$
(see the text).}
\end{figure}

Our discussion in this subsection so far has assumed $\omega_A=\omega_c$, which is 
a condition required for the BIC solution (Sec. III). A nonzero detuning $\delta \equiv \omega_c-\omega_A$ 
would lead to an incomplete destructive interference, and so photons would escape to the continuum
through the end cavities. However, we find that the rate of such a loss can be significantly 
suppressed in the $\chi \ll 1$ limit. Specifically, consider a system in a certain trapped state $|\beta_i \rangle$ with small
the magnitude of $\delta$ compared with $\lambda$, we then find that the probability loss rate $\gamma_d$ is proportional to 
$\gamma_c K(2M-K+1)\delta^2 \chi^2/\lambda^2$, which decreases with $\chi^2$ \cite{note2}.
Hence in the $\chi \ll 1$ limit the trapped states in the detuned system can be maintained 
in a relatively long (but finite) time scale.

\subsection{Linear analysis of the quantum-cavity regime} 

In this subsection we turn to the quantum-cavity
regime where $\chi  \gg 1$.
The fact that atoms are weakly excited in this regime allows us to 
make the following approximation:
\begin{equation}
J_{\mu}^{-} \approx \sqrt M d_{\mu}, \ \ \ J_{\mu}^{+} \approx  \sqrt M d_{\mu}^\dag,  \ \ \ J_{\mu}^{z}= d_{\mu}^\dag d_\mu -M/2,
\end{equation}
where $d_\mu$ and $d_{\mu} ^\dag$ ($\mu= L,R$) are annihilation and creation operators
satisfying the bosonic commutation relation $[d_{\mu},d_{\nu} ^\dag] = \delta_{\mu \nu}$. 
Such an approximation can be derived by keeping the leading term in the Holstein-Primakoff transformation \cite{HPTransformation},  which
is justified  because $\langle J_\mu ^ z\rangle \ll M/2$ for BICs in the $\chi \sqrt M \gg 1$ regime.

By using approximation (39), $H_1$ in Eq. (1) can be expressed as:
\begin{eqnarray}
H_1 & = & \omega_c(a^{\dagger}_La_L+a^{\dagger}_Ra_R +F_0^\dag F_0+F_+^\dag F_+ +F_-^\dag F_-) 
\nonumber\\
&& +  a^{\dagger}_{L} (\xi_+ F_+ + \xi_- F_-) + a_{L} (\xi_+ F_+^\dag +\xi_- F_-^\dag)  \nonumber \\
&& + a^{\dagger}_{R} (\xi_+F_+- \xi_-F_-) + a_R (\xi_+ F_+^\dag-\xi_- F_-^\dag)  ,
\end{eqnarray} 
where $\omega_A=\omega_c$ is assumed, and
\begin{eqnarray}
 && F_{+} = \frac{ {g \sqrt M (d_{L} + d_{R})+ 2 \lambda b_1}} {\sqrt{2} \sqrt{g^2 M +2  \lambda^2 } }   ,\\
&& F_{-} = \frac{ { d_{L} - d_{R}}} {\sqrt{2 } }  ,\\
&& F_{0} =  \frac{ g \sqrt M b_1 - \lambda  d_{L} - \lambda d_{R}} {\sqrt{  g^2 M +2 \lambda^2}} 
\end{eqnarray} 
are annihilation operators associated with three orthogonal polaritonic modes defined by the atomic oscillators and the middle cavity,
and they satisfy the commutation relations
$[F_{\alpha},F_{\alpha'}^\dag] = \delta_{\alpha \alpha'}$ for bosons. In addition, the coupling coefficients
$\xi_{\pm}$ are given by $\xi_+ =\sqrt{g^2 M + 2\lambda^2 }/ \sqrt{2}$ and $\xi_- = \sqrt{g^2 M}/\sqrt{2}$.

By Eq. (43), we see that the polaritonic mode $F_{0}$ does not couple to $a_R$ and $a_L$,
and hence $F_{0}$ is isolated from the continuous modes outside. The quantum cavity therefore refers to
the $F_0$ mode in which energy can be confined. In particular, the $F_0$ mode is mainly photonic because 
$F_0 \approx b_1$ in the $\chi \gg 1$ limit.

The performance of the quantum cavity is limited by various loss mechanisms. Here we 
examine the loss due to the spontaneous emission of photons from atoms into noncavity 
modes. Without loss of generality, a nonzero detuning $\delta \equiv \omega_c - \omega_A$ is included 
in the following discussion. Assuming atoms experience a collective decay, the damping can be introduced
by adding an imaginary part to the atomic frequency. Specifically, we make use of the linearized
Hamiltonian and replace
$\omega_A $  by $\omega_A - iM \gamma_A/2 $, where $\gamma_A$ is the single-atom spontaneous decay rate
(the factor $M$ is due to the collective spontaneous decay). Then by the 
Heisenberg's equations of motion, one can obtain a close set of differential equations for the expectation values of 
$a_{\mu} ,d_{\mu}$ and $b_1$:
\begin{eqnarray}
&& i \braket{\dot a_{\mu}} = \lambda \braket{b_1} + g \sqrt{M} \braket{d_{\mu}} + (\omega_c - i  \frac{\gamma_c}{2})\braket{a_{\mu}} ,\\
&& i \braket{\dot b_{1}} = \omega_c  \braket{b_1} +\lambda \braket{a_{L}} + \lambda \braket{a_{R}} ,\\
&& i \braket{\dot d_{\mu}} = g \sqrt{M} \braket{a_{\mu}} + (\omega_c -\delta -i \frac{M\gamma_A}{2})\braket{d_{\mu}},
\end{eqnarray}
where $\mu=L,R$ and the outside modes are assumed to be in the vacuum state.  Alternatively, one can obtain these equations from the 
master equation with collective atomic damping included.

We solve the eigenvalues of the linear system defined by the right hand side of Eqs. (44)-(46).
The eigenvalue with the smallest imaginary part (denoted as $\Gamma$) corresponds to the trapped 
mode, and $\Gamma$ is its decay rate. After some calculations, we have
 \begin{eqnarray}
\Gamma \approx \frac{M^2 \gamma_A  g^2+ \delta^2\gamma_c}{M^2g^4  +  \delta^2\gamma_c^2/4}\lambda^2,
\end{eqnarray} 
if $g \gg {\rm {max}}(\gamma_A, \gamma_c, \lambda)$. We characterize the performance of the quantum cavity
by the scaled quality factor $\tilde Q \equiv\gamma_c/\Gamma$, which is the (dimensionless) photon storage time of the quantum cavity 
in units of the free cavity decay time $\gamma_c^{-1}$. A quantum cavity with $\tilde Q \gg 1$ 
means that its photon storage time is much longer than that in the free cavity. 

\begin{figure}
\includegraphics[width=3 in]{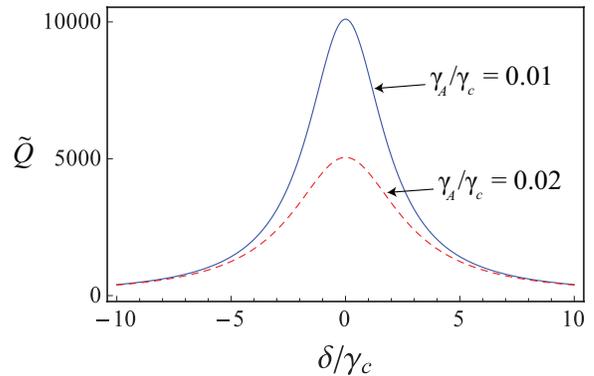}  
\caption{Scaled quality factor $\tilde Q =\gamma_c/\Gamma$  as a function of dimensionless detuning $\delta/\gamma_c$.
The parameters are $M=2$, $N=2$, $g/\lambda=10$, $\gamma_c/ \lambda = 1$.}
\end{figure}

In Fig. 4, we illustrate the dependence of $\tilde Q$ as a function of $\delta$. 
At resonance $\delta=0$, $\tilde Q = \chi^2 \gamma_c/\gamma_A$ is a maximum and it can be 
much larger than 1 in the $\chi \gg 1$ regime. In addition, 
by the width of the peak at half maximum, the high quality factor
can be maintained in a range of detunings when $ |\delta| < M g \sqrt {\gamma_A/ \gamma_c}$.

\section{Concluding Remarks}

To conclude we have obtained an analytic solution of a type of BIC formed by 
photons and atoms in a coupled-cavity array. These BICs 
originate from a destructive quantum interference effect 
which prevents excitations from being coupled to outside field modes, even though the energy lies 
in the continuous spectrum of the whole system. 
In addition, we identify a subradiance regime and a quantum-cavity regime in which most excitations 
are, respectively, atomic and photonic. 

We have also examined the triple-cavity configuration in detail.
In the subradiance regime we found that the dynamics is governed by the Tavis-Cummings Hamiltonian, 
which suggests that a mixture of nonradiating atomic states 
can be generated as a steady state of the free evolution problem. 
In the quantum-cavity regime, our analysis of
the linearized system, which includes atomic damping, 
has indicated that photons can be stored in the quantum cavity
with a life-time significantly longer than $\gamma_c^{-1}$. Such an efficient trapping of multiple photons 
may be used to explore quantum effects of photon-atom interactions and applications in the cavity QED \cite{rempe}
and circuit QED \cite{Koch}.

Finally, we would like to add two remarks on experimental aspects of our model. 
First, like many other BICs based on parameter-tuning \cite{review}, 
the existence of our BIC relies on the control of certain parameters, for example, $\omega_A=\Omega_q$. 
If these parameters are close but not exactly equal to the required values, a BIC would couple 
to the outside world and therefore has a finite life time (a leaky resonance). The dynamics can be analyzed 
by time-dependent perturbation theory, where the deviations of parameters are treated as a perturbation 
to the Hamiltonian. For a sufficiently weak perturbation, the system can still be trapped in 
the BIC with a time that can be long enough to produce observable effects. In other words, 
one may still study the BICs via leaky resonance in experiments without operating at 
perfect parameters. We have demonstrated this feature in Fig. 4, where the quantum cavity is 
able to store photons with a time much longer than the free cavity life time $1/\gamma_c$ 
even though the detuning is of the same size as the free cavity line width. 
Similarly, in the subradiance regime, the loss rate $\gamma_d$ due to a finite detuning 
can be significantly suppressed in the $\chi \ll 1$ limit.

Secondly, we note that decoherence of atoms is one of the main obstacles for experiments. 
For example, a spontaneous decay of an atom due to interaction with noncavity modes and
various dephasing mechanisms would kick the system out of the BIC. In particular, the quantum correlation between 
the two atomic ensembles, which keeps atoms from radiating, would be degraded. 
Therefore $\lambda$ and $g$
should be significantly larger than decoherence rates of the system so that effects of BICs (such as Fig. 3) can 
be established and observed before appreciable decoherence takes place. The recent parameters achievable in circuit QED 
are promising, because $\lambda$ and $g$ in the 10-100 MHz range can be much higher than decoherence rates 
(about 10-50 KHz) \cite{Koch}. In addition, the Tavis-Cummings model and its quantized energy spectrum have been demonstrated 
experimentally in circuit QED \cite{Fink}.


\begin{thebibliography}{99}
\bibitem {note1} It is worth noting that $\eta_L(\omega)$ and  
$\eta_R(\omega)$ can be different functions. Our
BICs, which are determined by the parameters of $H_1$ only, are independent of 
these functions. If the two ensembles of atoms are placed at cavities other than the two 
end cavities, i.e., any two positions in the array, the corresponding
Hamiltonian will take the same form, since two occupied cavities can be relabeled 
as end cavities. In this case the coupling to 
fields beyond the two occupied cavities will have a modified 
$\eta_L(\omega)$ and $\eta_R(\omega)$, but the form of BIC solution 
would be the same.
\bibitem{note2} The calculation is done by treating the detuning as a perturbation term in 
the Hamiltonian. By the first order time-dependent 
perturbation theory, the trapped state picks up a photon amplitude in end cavities
proportional to $\delta g \sqrt{K(2M-K+1)}/ \lambda^2$, and hence the photon probability 
in the end cavity scales as $\delta^2 g^2 K(2M-K+1)/ \lambda^4$. The probability loss rate 
is estimated by the product of this probability and the cavity field decay rate.
\end{thebibliography}
\end{document}